%% file: main.tex
\documentclass[10pt,conference]{IEEEtran}

\usepackage[english]{babel}
\usepackage[T1]{fontenc}
\usepackage[utf8x]{inputenc}

\usepackage{amsmath,amssymb,amsfonts}
\usepackage{graphicx}
\usepackage{enumerate}
\usepackage{listings}
\usepackage{hyperref}
\usepackage[font=small,skip=5pt]{caption}
\input{macros.tex}

\usepackage{xspace}
\usepackage{todonotes}

\newcommand{\snap}{{\small\textsf{Snap!}}\xspace}
\newcommand{\boogie}{{\small\textsf{Boogie}}\xspace}

\newcommand{\scratch}{{\small\textsf{Scratch}}\xspace}
\newcommand{\byob}{\text{BYOB}\xspace}
\newcommand{\DbC}{\textit{Design-by-Contract}\xspace}

\usepackage{xcolor}
\definecolor{codegreen}{rgb}{0,0.6,0}
\definecolor{codegray}{rgb}{0.5,0.5,0.5}
\definecolor{codepurple}{rgb}{0.58,0,0.82}
\definecolor{backcolour}{rgb}{0.95,0.95,0.92}

\usepackage{listings}
\usepackage{textcomp}

\lstdefinestyle{Boogie}{
    classoffset=0,
    morekeywords={var,while,procedure,forall},
    keywordstyle=\color{blue},
    classoffset=1,
    morekeywords={int},
    keywordstyle=\color{codegreen},
    classoffset=2,
    morekeywords={modifies,ensures,requires,invariant},
    keywordstyle=\color{codepurple},
    classoffset=0, % restore default
    backgroundcolor=\color{backcolour},
    commentstyle=\color{codegreen},
    numberstyle=\color{codegreen},
    basicstyle=\ttfamily\footnotesize,
    breaklines=true,
    captionpos=b,
    numbers=none
}

\begin{document}

\title{Teaching Design by Contract using Snap!
  \thanks{Identify applicable funding agency here. If none, delete this.}
}

\author{\IEEEauthorblockN{1\textsuperscript{st} Marieke Huisman}
\IEEEauthorblockA{\textit{Formal Methods and Tools} \\
\textit{University of Twente}\\
Enschede, The Netherlands\\
m.huisman@utwente.nl}
\and
\IEEEauthorblockN{2\textsuperscript{nd} Raúl E. Monti}
\IEEEauthorblockA{\textit{Formal Methods and Tools} \\
\textit{University of Twente}\\
Enschede, The Netherlands\\
r.e.monti@utwente.nl}
}
%-------------------------------------------------------------------------------

\maketitle

%-------------------------------------------------------------------------------

\begin{abstract}
  With the progress in deductive program verification research, new
  tools and techniques have become available to support
  design-by-contract reasoning about
  non-trivial programs written in widely-used programming
  languages. However, deductive program verification remains an
  activity for experts, with ample experience in programming,
  specification and verification. We would like to change this
  situation, by developing program verification techniques that are
  available to a larger audience. In this paper, we present
  how we developed prototypal program verification support for Snap!. Snap! is a
  visual programming language, aiming in particular at high school
  students. We added specification language constructs in a similar
  visual style, designed to make the intended semantics clear from the
  look and feel of the specification constructs. We provide support
  both for static and dynamic verification of Snap!  programs. Special
  attention is given to the error messaging, to make this as
  intuitive as possible. 
	% Finally, we outline how program verification
  % in Snap! could be introduced to high school students in a classroom
  % situation.
\end{abstract}


%-------------------------------------------------------------------------------

\input{sections/introduction.tex}
\input{sections/background.tex}
\input{sections/graphical_specification.tex}
\input{sections/error_reporting.tex}

\input{sections/tool_support.tex}
\input{sections/conclusions.tex}

{
\IEEEoverridecommandlockouts
\vspace*{-2\baselineskip}
\begin{IEEEbiography}
  [{\includegraphics[width=1in,height=1.25in,clip,keepaspectratio]{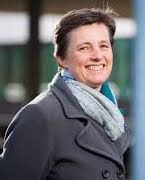}}]{Marieke
    Huisman} is a professor in Software Reliability at the University
  of Twente. She obtained her PhD in 2001 from the Radboud University
  Nijmegen. Afterwards she worked at INRIA Sophia Antipolis, and since
  2008 at University of Twente. Her research interests are in the
  verification of concurrent software, as implemented in the VerCors
  program verifier. She is in particular
  interested in making verification usable in a practical
  setting, and she works for example on annotation generation, and support for different programming languages. 
\end{IEEEbiography}
\vspace*{-2\baselineskip}
\begin{IEEEbiography} [{\includegraphics[width=1in,height=1.25in,clip,keepaspectratio]{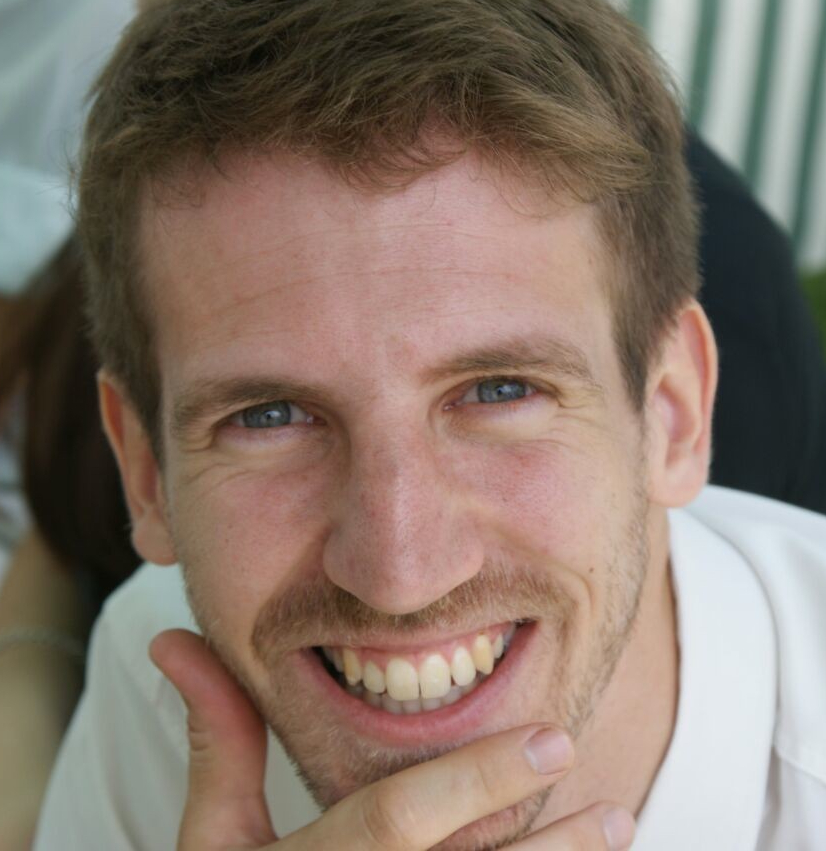}}]{Ra\'ul E. Monti}
received his PhD in 2018 from Universidad Nacional de C\'ordoba. He is currently a PostDoc at the Universiteit Twente.
His research interests involve the development and practical
application of formal foundations and tools for analysis and
verification of software and hardware systems by means of model
checking and deductive verification. His work involves interacting
with industry to apply his research in the verification of industrial
(embedded) systems and software.
\end{IEEEbiography}
}

%-------------------------------------------------------------------------------

\bibliographystyle{IEEEtran}
\bibliography{IEEEabrv,biblio}

\end{document}

%% file: macros.tex
\usepackage{xcolor}

\definecolor{lightblue}{RGB}{231,255,255}
\definecolor{lightred}{RGB}{255,224,224}
\definecolor{lightgreen}{RGB}{224,255,224}
\definecolor{lightyellow}{RGB}{255,255,224}
\definecolor{lightpurple}{RGB}{255,224,255}
\definecolor{darkerred}{RGB}{64,0,0}
\definecolor{darkred}{RGB}{128,0,0}
\definecolor{darkblue}{RGB}{0,0,128}
\definecolor{darkgreen}{RGB}{0,128,0}
\definecolor{darkpurple}{RGB}{128,0,128}
\definecolor{black}{RGB}{0,0,0}

\makeatletter
\def\THICKhrulefill{\leavevmode \leaders \hrule height 5pt\hfill \kern \z@}
\makeatother

%% file: sections/introduction.tex
\section{Introduction}

Research in deductive program verification has made substantial progress over the last years: tools and technique have been developed to reason about non-trivial programs written in widely-used programming languages, the level of automation has substantially increased, and bugs in widely-used libraries have been found \cite{DeGouwRBBH15,oortwijn2020automated,SOJH2020}. However, the use of deductive verification techniques remains the field of expert users, and substantial programming knowledge is necessary to appreciate the benefits of these techniques.

% We feel that it is important to change this situation, and to make deductive program verification techniques accessible to novice programmers. Specifying the intended behaviour of a program explicitly is something that programmers should learn about from the beginning, and should be an integral part of the process leading from design to implementation. Therefore, we feel that it is important that the \DbC approach~\cite{DBLP:journals/computer/Meyer92} (DbC), which lies at the core of deductive program verification is taught in first year Computer Science curricula. In this paper, we take this even further, and make the \DbC idea accessible to high school students, in combination with appropriate tool support.

We believe that it is important to make deductive program verification techniques accessible also to novice programmers. Therefore, we have to teach the \DbC~\cite{DBLP:journals/computer/Meyer92} (DbC) approach, which requires the programmer to explicitly specify the assumptions and responsibilities of code in a modular way, in parallel with actually teaching programming, i.e. DbC should be taught as an integral part of the process leading from design to implementation. In this paper, we make the \DbC idea accessible to high school students, in combination with appropriate tool support, which is currently
unavailable.

Concretely, this paper presents a \DbC approach for \snap~\cite{harvey2013snap}. \snap is a visual programming language targeting high school students. The design of \snap is inspired by \scratch, another widely-used visual programming language. Compared to \scratch, \snap has some more advanced programming features. In particular, \snap provides the possibility to create parametrised reusable blocks, basically modelling user-defined functions. Also the look and feel of \snap targets high school students, whereas \scratch aims at an even younger age group. \snap has been successfully integrated in high school curricula, by its integration in the \emph{Beauty and Joy of Computing} course~\cite{DBLP:journals/inroads/GarciaHB15}. This course combines programming skills with a training in abstract computational thinking. 

The first step to support \DbC for \snap is to define a suitable specification language. The visual specification language that we propose in this paper is built as a seamless extension of \snap, i.e. we propose a number of new specification blocks and natural modifications of existing ones. These variations capture the main ingredients for the \DbC approach, such as pre- and postconditions. Moreover, we also provide blocks to add assertions and loop invariants in a program and we extend the standard expression pallets of \snap with some common expressions to ease specifications. The choice of specification constructs is inspired by existing specification languages for \DbC, such as JML~\cite{Leavens99}, choosing the most frequently used constructs with a clear and intuitive meaning. Moreover, all verification blocks are carefully designed to reflect the intended semantics of the specifications in a visual way. 

A main concern for a programmer, after writing the specification of the intended behaviour of their programs, should be to validate that these programs behave according to their specification. Therefore, we provide two kinds of tool support: (i)~runtime assertion checking~\cite{Cheon03}, which checks whether specifications are not violated during a particular program execution, and (ii)~static checking (or deductive verification)~\cite{leino1995towards}, which verifies that all possible program executions respect its specifications. The runtime assertion checker is built as an extension of the standard \snap execution mechanism. The deductive verification support is built by providing a translation from a \snap program into \boogie~\cite{DBLP:conf/fmco/BarnettCDJL05}. 

Another important aspect to take into account for a good learning experience are the error messages that indicate that a specification is violated. We have integrated these messages in \snap's standard error reporting system, again sticking to the look and feel of standard \snap. Moreover, we have put in effort to make the error messages as clear as possible, so that also a relatively novice programmer can understand why the implementation deviates from the specification.

%% file: sections/background.tex
\section{Background}\label{sec:background}

%--------------------------------------
\subsection{\snap}

\snap\ is a visual programming language. It has been designed to introduce children (but also adults) to programming in an intuitive way. At the same time, it is also a platform for serious study of computer science\cite{harvey2017snap}. \snap actually re-implements and extends \scratch~\cite{DBLP:journals/cacm/ResnickMMREBMRSSK09}. Programming in \snap is done by  dragging and dropping blocks into the coding area. Blocks represent common program constructs such as variable declarations, control flow statements (branching and loops), function calls and assignments. Snapping blocks together, the user builds a script and visualises its behaviour by means of turtle graphics visualisations, called \textit{sprites}. Sprites can change shape, move, show bubbled text, play music, etc. For all these effects, dedicated blocks are available.

\begin{figure}[h]
\includegraphics[width=\linewidth]{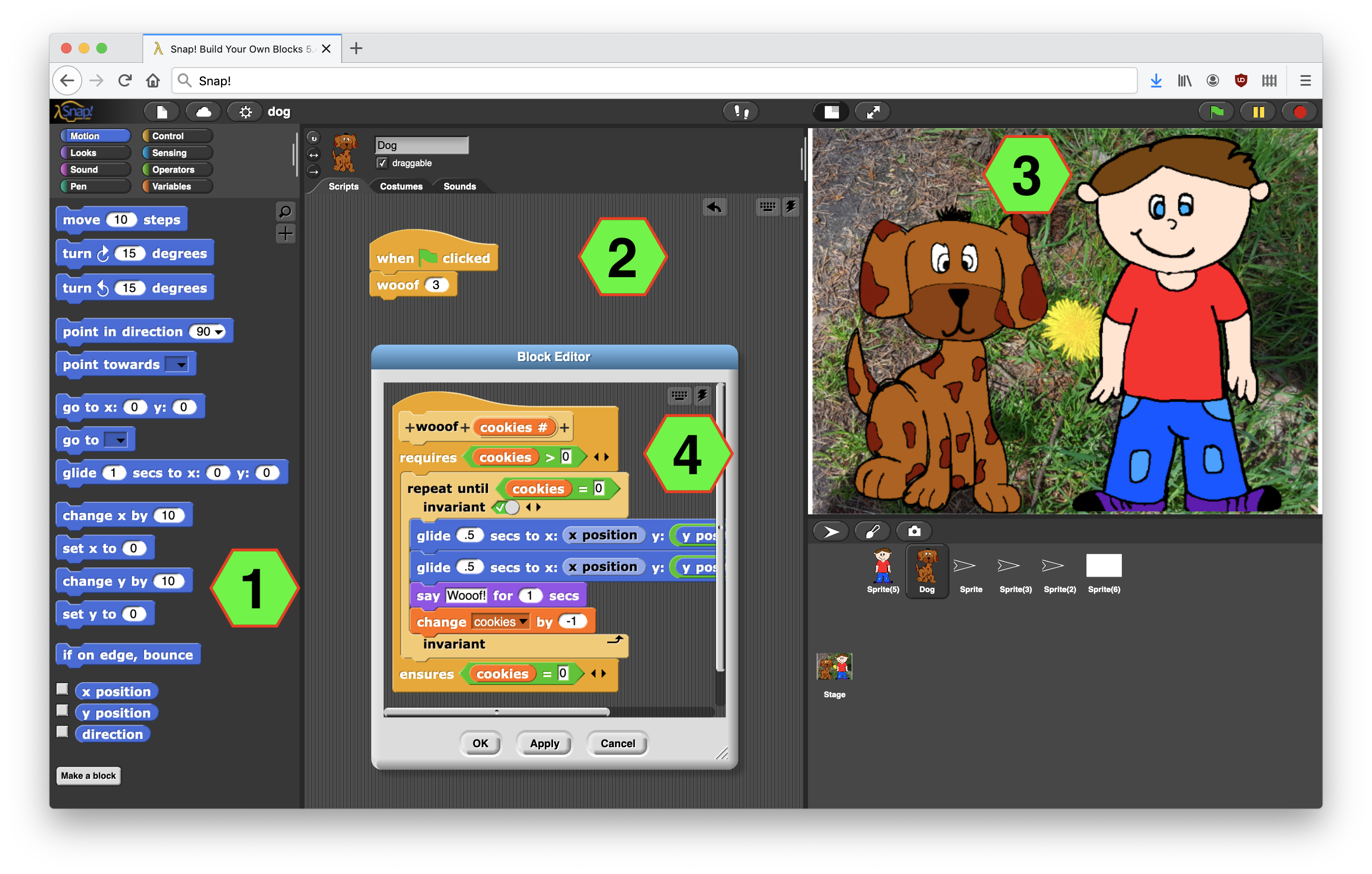}
\caption{The \snap working area.}
\label{fig:workingarea}
\end{figure}

The \snap\ interface divides the working area into three parts: the pallet area, the scripting area, and the stage area, indicated by labels 1, 2 and 3, respectively, in Fig.~\ref{fig:workingarea}.
On the left, the various programming blocks are organised into pallets that describe their natural use. For instance, the \emph{Variables} pallet contains blocks for declaring and manipulating variables. In \snap, variables are dynamically typed.
Blocks are dragged and dropped from the pallets into the scripting area, located at the centre of the working area where the \snap program is constructed.
Blocks can be arranged by snapping them together, or by inserting them as arguments of other blocks. Blocks can only be used as arguments if their shapes match with the shape of the argument slots in the target block. These shapes actually provide a hint on the expected evaluation type of a block, 
for instance, rounded slots for numbers $\vcenter{\hbox{\includegraphics[height=\baselineskip]{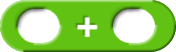}}}$ and diamond slots for booleans $\vcenter{\hbox{\includegraphics[height=\baselineskip]{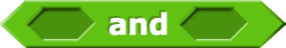}}}$.

The behaviour of the script is shown with turtle graphics drawings in the stage area located in the rightmost part of the screen.

In addition, at the bottom of the pallet area, there is a ``Make a block'' button. This allows the user to define his or her \emph{Build Your Own Block} (\byob) blocks. When pressed, a new floating ``Block Editor'' window pops out with a new coding area, in which the behaviour of the personalised block can be defined (similar to how a script is made in the scripting area). Label $4$ in Fig.~\ref{fig:workingarea} shows a \byob block being edited. Once defined, the \byob block becomes available to be used just as any other predefined block.

%--------------------------------------
\subsection{Program Verification}

The basis of the \DbC approach~\cite{meyer1987eiffel} is that the behaviour of all program components is defined as a contract. For example, a function contract specifies the conditions under which a function may be called (the function's \emph{precondition}), and it specifies the guarantees that the function provides to its caller (the function's \emph{postcondition}). There exist several specification languages that have their roots in this \DbC approach. For example the Eiffel programming language has built-in support for pre- and postconditions~\cite{meyer1988eiffel}, and for Java, the behavioural interface language JML~\cite{leavens2005design} is widely used. As is common for such languages, we use the keyword \emph{requires} to indicate a precondition, and the keyword \emph{ensures} to indicate a postcondition.

If a program behaviour is specified using contracts, various techniques can be used to validate whether an implementation respects the contract.

Dynamic verification validates an implementation w.r.t. a specification at runtime. This means that, whenever during program execution a specification is reached, it will be checked for this particular execution that the property specified indeed holds. In particular, this means that whenever a function will be called, its precondition will be checked, and whenever the function returns, its postcondition will be checked.
An advantage of this approach is that it is easy and fast to use it: one just runs a program and checks if the execution does not violate the specifications. A disadvantage is that it only provides guarantees about a concrete execution.

In contrast, static verification aims at verifying that all possible behaviours of a function respect its contract. This is done by applying Hoare logic proof rules~\cite{Hoare69} or using Dijkstra's predicate transformer semantics~\cite{Dijkstra76}. Applying these rules results in a set of first-order proof obligations; if these proof obligations can be proven it means that the code satisfies its specification. Advantage of this approach is that it guarantees correctness of all possible behaviours. Disadvantage is that it is often labour-intensive, and often many additional annotations, such as for example loop invariants, are needed to guide the prover.

%% file: sections/graphical_specification.tex
\section{Visual Program Specifications}
\label{sec:visual}

This section discusses how to add visual specification constructs to \snap. Our goal was to do this in such a way that (1)~the intended semantics of the specification construct is clear from the way it is visualised, and (2)~that it smoothly integrates with the existing programming constructs in \snap

Often, Design-by-Contract specifications are added as special comments in the code. For example, in JML a function contract is written in a special comment, tagged with an \texttt{@}-symbol, immediately preceding the function declaration. The tag ensures that the comment can be recognised as part of the specification. There also exist languages where for example pre- and postconditions are part of the language (e.g., Eiffel~\cite{Meyer91}, Spec\#~\cite{BarnettLW04}). We felt that for our goal, specifications should be integrated in a natural way in the language, rather than using comments. 
Therefore, we introduce variations of the existing block structures, in which we added suitable slots for the specifications. This section discusses how we added pre- and postconditions, and in-code specifications such as asserts and loop invariants to \snap. In addition, to have a sufficiently expressive property specification language, we also propose an extension of the expression constructs. 

%--------------------------------------

\subsection{Visual Pre- and Postconditions}

\begin{figure}[htb]
	\centering
	\includegraphics[width=.95\linewidth]{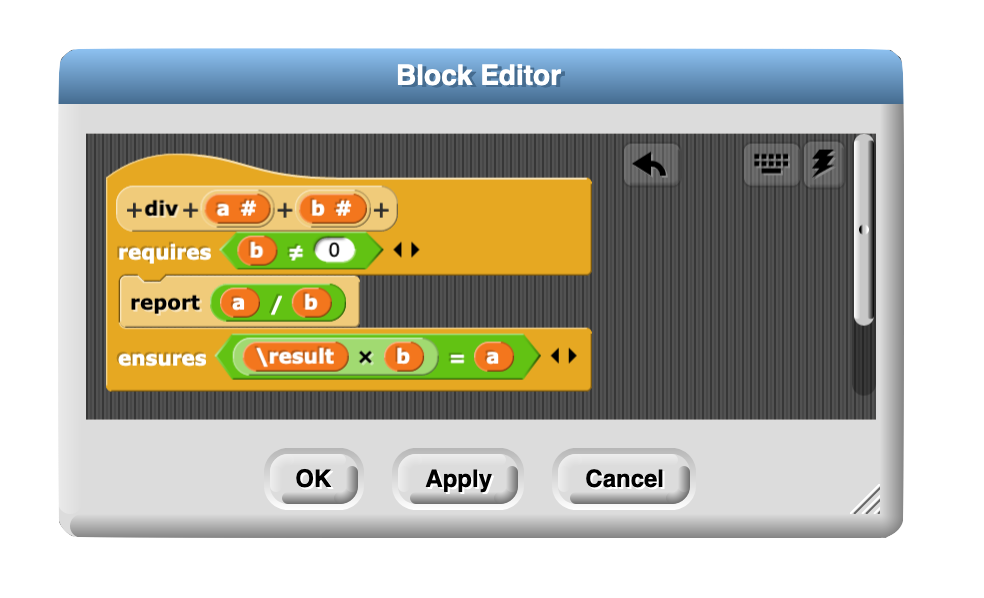}
	\vspace{-5pt}
	\caption{Hat block extended with contracts}
	\label{fig:contracts3}
\end{figure}

To specify pre- and postconditions for a \byob script, we provide
a variation of the initial hat block with a slot for a precondition at the start of the block, and a slot for a postcondition at the end of the block (Fig.~\ref{fig:contracts3}).

This shape is inspired by the c-shaped style of other \snap blocks, such as blocks for loops. The main advantage is that it visualises  at which points in the execution, the pre- and the postconditions are expected to hold. In addition, it also graphically identifies which code is actually verified. Moreover, the shapes are already familiar to the \snap programmer. If the slots are not filled, default pre- and postcondition \texttt{true} can be used. 
Notice that the pre- and postcondition slots consist of multiple boolean-argument slots, and we define the property to be the conjunction of the evaluation of each of these slots. 
This is similar to how \snap extends a list or adds arguments to the header of a \byob.

\vspace{-10pt}
\vspace{-5pt}

\subsection{Visual Assertions and Loop Invariants}

For static verification, pre- and postconditions are often not sufficient, and we need additional in-code specifications to guide the prover, such as assertions, which specify properties that should hold at a particular point in the program, and loop invariants. Moreover, assertions can also be convenient for run-time assertion checking to make it explicit that a property holds at a particular point in the program.

\paragraph{Visual Assertions}
To specify assertions, both the property specified and the location within the code are relevant. To allow the specification of assertions at arbitrary places in a script, we define a special assertion block $\vcenter{\hbox{\includegraphics[height=\baselineskip]{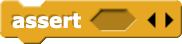}}}$ similar to all other control blocks.

\begin{figure}[htb]
\centering
\includegraphics[width=.55\linewidth]{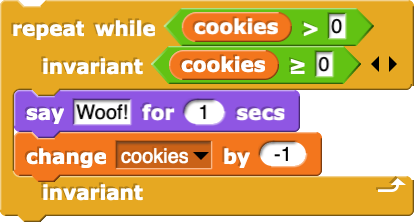}
\vspace{5pt}
\caption{Visual loop invariants.}
\vspace{-10pt}
\label{fig:loopinvariant}
\end{figure}

\paragraph{Visual Loop Invariants}
Loop invariants are necessary for static verification~\cite{Tuerk10}. A loop invariant should hold at the beginning and end of every loop iteration. 
To account for this, we provide a (multi-argument boolean) slot to specify the loop invariant in the traditional \snap c-shaped loop block.
This slot is located just after the header where the loop conditions are defined. In addition, the c-shaped loop block repeats the word invariant at the bottom of the block (see Figure~\ref{fig:loopinvariant}) to visually indicate that the invariant is checked after each iteration. 

\subsection{Visual Expressions}

In addition, we have introduced some specification-only keywords, as commonly found in Design-by-Contract languages. 

\begin{itemize}
  \item An \emph{old} expression is used in postconditions to indicate that a variable/expression should be evaluated in the pre-state of the function. To support this, we introduced an operator block $\vcenter{\hbox{\includegraphics[height=\baselineskip]{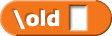}}}$ with a slot for a variable name.
  \item A \emph{result} expression refers to the return value of a function inside its postcondition. We support this by introducing a constant $\vcenter{\hbox{\includegraphics[height=\baselineskip]{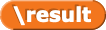}}}$ operator, that allows to specify a property about the result value of a reporter \byob. 
\end{itemize}

\begin{figure}[htb]
\includegraphics[height=1.45em]{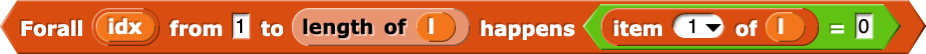}
\caption{A global quantification expression block}
\label{fig:quantifiers}
\end{figure}

We also introduce syntax to ease the definition of complex Boolean expressions, by means of the operator blocks
$\vcenter{\hbox{\includegraphics[height=\baselineskip]{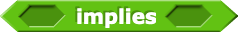}}}$,
$\vcenter{\hbox{\includegraphics[height=\baselineskip]{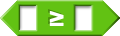}}}$,
$\vcenter{\hbox{\includegraphics[height=\baselineskip]{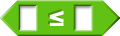}}}$ and
$\vcenter{\hbox{\includegraphics[height=\baselineskip]{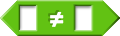}}}$, as well as syntax to write more advanced Boolean expressions, introducing support for quantified expressions (See Fig.\ref{fig:quantifiers}).

%% file: sections/error_reporting.tex
\section{Graphical approach to verification result reporting}
\label{sec:report}

Another important point to consider is how to report on the outcome of the verification: (1)~presenting the verdict of a passed verification, and (2)~in case of failure, giving a concrete and understandable explanation for the failure. The latter is especially important in our case, as we are using the technique with inexperienced users. 

In order to signal a contract violation, or any assertion invalidated during dynamic verification, we use \snap's pop-up notification windows. These windows have the advantage that a failing block can be printed inside them even when the failing script is not currently visible to the user. This allows to be very precise about the error, even  when the \byob body is not currently visible.

\begin{figure}
\centering
\includegraphics[width=.95\linewidth]{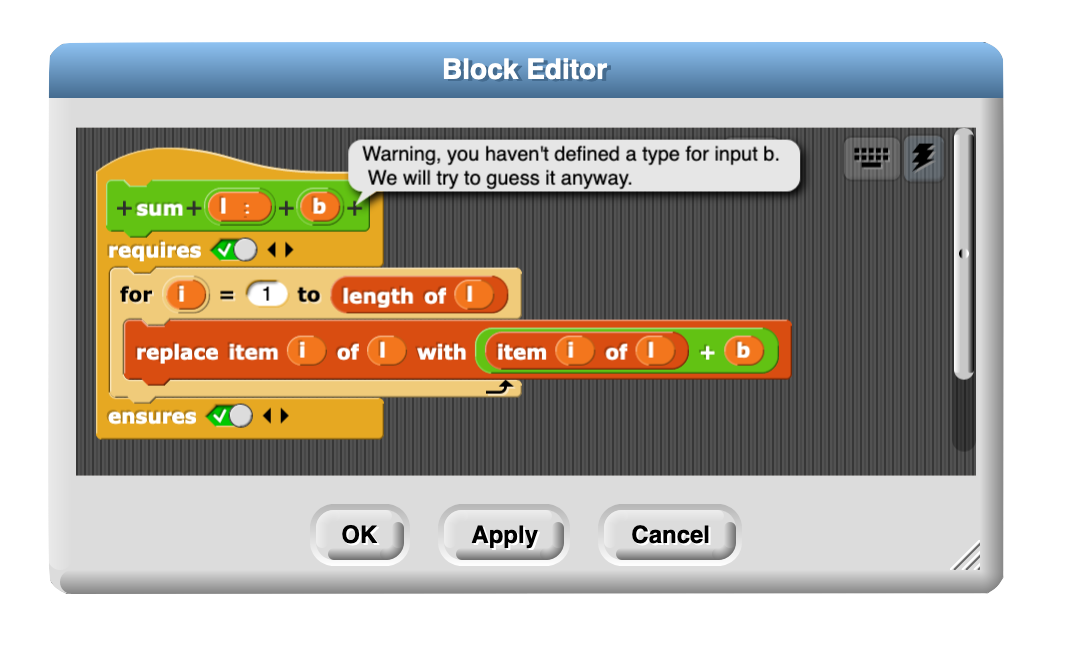}
\vspace{-5pt}
\caption{Static verification compilation notification.}
\label{fig:bubbles}
\end{figure}

In order to signal errors while compiling to \boogie, such as making use of dynamic typing or nested lists in your \snap \byob code, we use \snap's speech bubbles that can emerge at specific points in the script while describing the cause of failure. This has the advantage that the failing block can easily be singled out by the location of the bubble, while the cause of failure is described by the text inside the bubble. We find this option less invasive than a pop-up window but still as precise, and we can be sure that the blocks involved will be visible since static verification is triggered from the \byob editor window (See Fig.\ref{fig:bubbles}).
Notice that currently we do not report the results of static verification within \snap, since our extension only returns a compiled \boogie code which has to be verified with \boogie separately.

%% file: sections/tool_support.tex
\section{Tool support}
\label{sec:toolsupport}

We have developed our ideas into a prototypal extension to \snap\ which can be found at \url{https://git.snt.utwente.nl/montire/verifiedsnap/}. This repository
also contains a set of running examples to showcase the new support for verification. These are available in the \emph{lessons} folder under the root directory along with an exercise sheet named \emph{exercises.pdf}. The extension uses the same technology as the original \snap\ and can be run by just opening the \emph{snap.html} file in most common web-browsers that support java-script. 

Our extension supports both dynamic and static verification of \byob blocks. Dynamic verification is automatically triggered when executing \byob blocks in the usual way. For static verification, a dedicated button located at the top right corner of the \byob editor window allows to trigger the compilation of the \byob code into an intended equivalent \boogie code. The compiled code can be then downloaded and verified with \boogie. \boogie can be run locally or on the cloud at \url{https://rise4fun.com/Boogie/}.
\emph{Dynamic verification} has been fully integrated into the normal execution flow of a \snap\ program, and thus there is no real restrictions on the characteristics of the \byob that can be dynamically verified. 
For \emph{Static verification}, we have restricted data types to be Integers, Booleans and List of Integers. Moreover, we do not support dynamic typing of variables. 
Finally, we only focus on compiling an interesting subset of \snap blocks for the sake of teaching \DbC.

%% file: sections/conclusions.tex
\section{Conclusions}
\label{sec:conclusion}

This paper presented a prototypal program verification extension to  \snap. The extension is intended to support the teaching of \DbC in the later years of high schools. We paid considerable attention to the didactic aspects of our tool: the looks and feel of the extension should remain familiar to \snap users, the syntax and structure of the new blocks should give a clear intuition about their semantics, and the error reporting should be precise and expressive.
  
  Our extension allows to analyse \byob blocks both by runtime assertion checking and static verification. Runtime assertion checking is fully integrated into \snap and there is no limitation on the kind of blocks that can be analysed. Static verification compiles the \snap code into a \boogie equivalent code and the verification needs to be run outside of \snap. Moreover, we make some restrictions on the kind of \byob blocks we can compile, in order to keep the complexity of the prototype low. As future work we would like to lift these restrictions as much as possible by integrating the remaining \snap blocks into the compilation and by allowing other data types to be used. Also, we would like to integrate the verification into \snap, translating \boogie messages back to the \snap world, to help student to interpret them.

 We would like to carry out an empirical study on our proposed approach. This will require the development of a concrete study plan and its evaluation in a Dutch classroom.

  Computer science curricula that uses blocks programming is widely and freely available \cite{BJCGuide,GoogleCSFirst,CCC,PencilCodeManual,factorovich2015actividades}. Nevertheless, it is hardly spotted that they include topics around design and verification of code. The words `test' or `testing' are also rare around the curricula and, where mentioned, they are not sufficiently motivated. The drawbacks of teaching coding with blocks without paying attention to design nor correctness has already been analysed \cite{DBLP:conf/iticse/Meerbaum-SalantAB11,aivaloglou2016kids}. We have not found any work on teaching these concepts in schools, nor implementations on block programming that support teaching them.

%% file: main.bbl
\begin{thebibliography}{10}
\providecommand{\url}[1]{#1}
\csname url@samestyle\endcsname
\providecommand{\newblock}{\relax}
\providecommand{\bibinfo}[2]{#2}
\providecommand{\BIBentrySTDinterwordspacing}{\spaceskip=0pt\relax}
\providecommand{\BIBentryALTinterwordstretchfactor}{4}
\providecommand{\BIBentryALTinterwordspacing}{\spaceskip=\fontdimen2\font plus
\BIBentryALTinterwordstretchfactor\fontdimen3\font minus
  \fontdimen4\font\relax}
\providecommand{\BIBforeignlanguage}[2]{{%
\expandafter\ifx\csname l@#1\endcsname\relax
\typeout{** WARNING: IEEEtran.bst: No hyphenation pattern has been}%
\typeout{** loaded for the language `#1'. Using the pattern for}%
\typeout{** the default language instead.}%
\else
\language=\csname l@#1\endcsname
\fi
#2}}
\providecommand{\BIBdecl}{\relax}
\BIBdecl

\bibitem{DeGouwRBBH15}
S.~De~Gouw, J.~Rot, F.~De~Boer, R.~Bubel, and R.~H{\"a}hnle, ``{OpenJDK's}
  java.utils.collection.sort() is broken: The good, the bad and the worst
  case,'' in \emph{Proc.\ 27th Intl.\ Conf.\ on Computer Aided Verification
  (CAV), San Francisco}, ser. LNCS, D.~Kroening and C.~Pasareanu, Eds., vol.
  9206.\hskip 1em plus 0.5em minus 0.4em\relax Springer, Jul. 2015, pp.
  273--289.

\bibitem{oortwijn2020automated}
W.~Oortwijn, M.~Huisman, S.~J. Joosten, and J.~van~de Pol, ``Automated
  verification of parallel nested dfs,'' in \emph{International Conference on
  Tools and Algorithms for the Construction and Analysis of Systems}.\hskip 1em
  plus 0.5em minus 0.4em\relax Springer, 2020, pp. 247--265.

\bibitem{SOJH2020}
\BIBentryALTinterwordspacing
M.~Safari, W.~Oortwijn, S.~Joosten, and M.~Huisman, ``Formal verification of
  parallel prefix sum,'' in \emph{NASA Formal Methods}, R.~Lee, S.~Jha, and
  A.~Mavridou, Eds.\hskip 1em plus 0.5em minus 0.4em\relax Cham: Springer
  International Publishing, 2020, pp. 170--186. [Online]. Available:
  \url{https://doi.org/10.1007/978-3-030-55754-6_10}
\BIBentrySTDinterwordspacing

\bibitem{DBLP:journals/computer/Meyer92}
\BIBentryALTinterwordspacing
B.~Meyer, ``Applying "design by contract",'' \emph{Computer}, vol.~25, no.~10,
  pp. 40--51, 1992. [Online]. Available: \url{https://doi.org/10.1109/2.161279}
\BIBentrySTDinterwordspacing

\bibitem{harvey2013snap}
B.~Harvey, D.~D. Garcia, T.~Barnes, N.~Titterton, D.~Armendariz, L.~Segars,
  E.~Lemon, S.~Morris, and J.~Paley, ``Snap!(build your own blocks),'' in
  \emph{Proceeding of the 44th ACM technical symposium on Computer science
  education}, 2013, pp. 759--759.

\bibitem{DBLP:journals/inroads/GarciaHB15}
\BIBentryALTinterwordspacing
D.~Garcia, B.~Harvey, and T.~Barnes, ``The beauty and joy of computing,''
  \emph{Inroads}, vol.~6, no.~4, pp. 71--79, 2015. [Online]. Available:
  \url{https://doi.org/10.1145/2835184}
\BIBentrySTDinterwordspacing

\bibitem{Leavens99}
G.~T. Leavens, A.~L. Baker, and C.~Ruby, ``{JML}: A notation for detailed
  design,'' in \emph{Behavioral Specifications of Businesses and Systems},
  H.~Kilov, B.~Rumpe, and I.~Simmonds, Eds.\hskip 1em plus 0.5em minus
  0.4em\relax Boston, MA: Springer US, 1999, pp. 175--188.

\bibitem{Cheon03}
Y.~Cheon, ``A runtime assertion checker for the {Java} {M}odeling {L}anguage,''
  Ph.D. dissertation, Department of Computer Science, Iowa State University,
  Ames, 2003, technical Report 03-09.

\bibitem{leino1995towards}
K.~R.~M. Leino, ``Towards reliable modular programs,'' 1995.

\bibitem{DBLP:conf/fmco/BarnettCDJL05}
\BIBentryALTinterwordspacing
M.~Barnett, B.~E. Chang, R.~DeLine, B.~Jacobs, and K.~R.~M. Leino, ``Boogie:
  {A} modular reusable verifier for object-oriented programs,'' in \emph{Formal
  Methods for Components and Objects, 4th International Symposium, {FMCO} 2005,
  Amsterdam, The Netherlands, November 1-4, 2005, Revised Lectures}, ser.
  Lecture Notes in Computer Science, F.~S. de~Boer, M.~M. Bonsangue, S.~Graf,
  and W.~P. de~Roever, Eds., vol. 4111.\hskip 1em plus 0.5em minus 0.4em\relax
  Springer, 2005, pp. 364--387. [Online]. Available:
  \url{https://doi.org/10.1007/11804192\_17}
\BIBentrySTDinterwordspacing

\bibitem{harvey2017snap}
B.~Harvey and J.~M{\"o}nig, ``Snap! reference manual,'' \emph{URL http://snap.
  berkeley. edu/SnapManual. pdf}, 2017.

\bibitem{DBLP:journals/cacm/ResnickMMREBMRSSK09}
\BIBentryALTinterwordspacing
M.~Resnick, J.~Maloney, A.~Monroy{-}Hern{\'{a}}ndez, N.~Rusk, E.~Eastmond,
  K.~Brennan, A.~Millner, E.~Rosenbaum, J.~S. Silver, B.~Silverman, and Y.~B.
  Kafai, ``Scratch: programming for all,'' \emph{Commun. {ACM}}, vol.~52,
  no.~11, pp. 60--67, 2009. [Online]. Available:
  \url{https://doi.org/10.1145/1592761.1592779}
\BIBentrySTDinterwordspacing

\bibitem{meyer1987eiffel}
B.~Meyer, J.-M. Nerson, and M.~Matsuo, ``Eiffel: object-oriented design for
  software engineering,'' in \emph{European Software Engineering
  Conference}.\hskip 1em plus 0.5em minus 0.4em\relax Springer, 1987, pp.
  221--229.

\bibitem{meyer1988eiffel}
B.~Meyer, ``Eiffel: A language and environment for software engineering,''
  \emph{Journal of Systems and Software}, vol.~8, no.~3, pp. 199--246, 1988.

\bibitem{leavens2005design}
G.~T. Leavens, Y.~Cheon, C.~Clifton, C.~Ruby, and D.~R. Cok, ``How the design
  of jml accommodates both runtime assertion checking and formal
  verification,'' \emph{Science of Computer Programming}, vol.~55, no. 1-3, pp.
  185--208, 2005.

\bibitem{Hoare69}
C.~Hoare, ``An axiomatic basis for computer programming,'' \emph{Communications
  of the ACM}, vol.~12, no.~10, pp. 576--580, 1969.

\bibitem{Dijkstra76}
E.~Dijkstra, \emph{A Discipline of Programming}.\hskip 1em plus 0.5em minus
  0.4em\relax Prentice-Hall, 1976.

\bibitem{Meyer91}
\BIBentryALTinterwordspacing
B.~Meyer, \emph{Eiffel: The Language}.\hskip 1em plus 0.5em minus 0.4em\relax
  Prentice-Hall, 1991. [Online]. Available:
  \url{http://www.eiffel.com/doc/\#etl}
\BIBentrySTDinterwordspacing

\bibitem{BarnettLW04}
M.~Barnett, K.~R.~M. Leino, and W.~Schulte, ``The {Spec\#} programming system:
  An overview,'' in \emph{Construction and Analysis of Safe, Secure and
  Interoperable Smart Devices: Proceedings of the International Workshop
  CASSIS~2004}, ser. LNCS, G.~Barthe, L.~Burdy, M.~Huisman, J.-L. Lanet, and
  T.~Muntean, Eds., vol. 3362.\hskip 1em plus 0.5em minus 0.4em\relax Springer,
  2005, pp. 151--171.

\bibitem{Tuerk10}
T.~T\"{u}rk, ``Local reasoning about while-loops,'' in \emph{VSTTE 2010.
  Workshop Proceedings}, R.~Joshi, T.~Margaria, P.~M{\"u}ller, D.~Naumann, and
  H.~Yang, Eds.\hskip 1em plus 0.5em minus 0.4em\relax ETH Z\"{u}rich, 2010,
  pp. 29--39.

\bibitem{BJCGuide}
``{The Beauty and Joy of Computing. An AP CS Principles Course},''
  \url{https://bjc.edc.org/}. Accessed October 2020.

\bibitem{GoogleCSFirst}
``{CS First},'' \url{https://csfirst.withgoogle.com/s/en/home}. Accessed
  October 2020.

\bibitem{CCC}
``{The Creative Computing Curriculum},''
  \url{http://creativecomputing.gse.harvard.edu/guide/}. Accessed October 2020.

\bibitem{PencilCodeManual}
``{An Introduction to Programming. A Pencil Code Teacher's Manual},''
  \url{https://manual.pencilcode.net/}. Accessed October 2020.

\bibitem{factorovich2015actividades}
P.~Factorovich and F.~Sawady, \emph{Actividades para aprender a Program. AR:
  Segundo ciclo de la educaci{\'o}n primaria y primero de la secundaria}.\hskip
  1em plus 0.5em minus 0.4em\relax Miller Ed Buenos Aires, 2015.

\bibitem{DBLP:conf/iticse/Meerbaum-SalantAB11}
\BIBentryALTinterwordspacing
O.~Meerbaum{-}Salant, M.~Armoni, and M.~Ben{-}Ari, ``Habits of programming in
  scratch,'' in \emph{Proceedings of the 16th Annual {SIGCSE} Conference on
  Innovation and Technology in Computer Science Education, ITiCSE 2011,
  Darmstadt, Germany, June 27-29, 2011}, G.~R{\"{o}}{\ss}ling, T.~L. Naps, and
  C.~Spannagel, Eds.\hskip 1em plus 0.5em minus 0.4em\relax {ACM}, 2011, pp.
  168--172. [Online]. Available: \url{https://doi.org/10.1145/1999747.1999796}
\BIBentrySTDinterwordspacing

\bibitem{aivaloglou2016kids}
E.~Aivaloglou and F.~Hermans, ``How kids code and how we know: An exploratory
  study on the scratch repository,'' in \emph{Proceedings of the 2016 ACM
  Conference on International Computing Education Research}, 2016, pp. 53--61.

\end{thebibliography}
